\documentclass[tickz,longauth]{aa}
\usepackage[dvipsnames]{xcolor}
\usepackage{url,hyperref}
\usepackage{tikz}
\usepackage[switch]{lineno}
\hypersetup{
    colorlinks=true,
    linkcolor=Lavender,
    citecolor=PineGreen,       
    filecolor=magenta,      
    urlcolor=cyan           
}
\usepackage{graphicx}
\usepackage{amsmath}
\usepackage{mathabx}
\usepackage{txfonts}

\usepackage{natbib,twoopt}
\bibpunct{(}{)}{;}{a}{}{,}             
\makeatletter
  \newcommandtwoopt{\citeads}[3][][]{\href{http://adsabs.harvard.edu/abs/#3}%
    {\def\hyper@linkstart##1##2{}%
     \let\hyper@linkend\@empty\citealp[#1][#2]{#3}}}
  \newcommandtwoopt{\citepads}[3][][]{\href{http://adsabs.harvard.edu/abs/#3}%
    {\def\hyper@linkstart##1##2{}%
     \let\hyper@linkend\@empty\citep[#1][#2]{#3}}}
  \newcommandtwoopt{\citetads}[3][][]{\href{http://adsabs.harvard.edu/abs/#3}%
    {\def\hyper@linkstart##1##2{}%
     \let\hyper@linkend\@empty\citet[#1][#2]{#3}}}
  \newcommandtwoopt{\citeyearads}[3][][]%
    {\href{http://adsabs.harvard.edu/abs/#3}
    {\def\hyper@linkstart##1##2{}%
     \let\hyper@linkend\@empty\citeyear[#1][#2]{#3}}}
    \makeatother

\newcommand{\hess}{H.E.S.S.}
\newcommand{\swift}{\textit{Swift}}
\newcommand{\xrt}{\textit{Swift}-XRT}
\newcommand{\uvot}{\textit{Swift}-UVOT}
\newcommand{\fermilat}{\textit{Fermi}-LAT}
\newcommand{\pks}{PKS~0736+017}
\newcommand{\gray}{$\gamma$-ray}
\newcommand{\graynohy}{$\gamma$ ray}

\makeatletter
\renewcommand*{\@fnsymbol}[1]{\ifcase#1\or*\or$\dagger$\or$\ddagger$\or**\or$\dagger\dagger$\or$\ddagger\ddagger$\fi}
\makeatother

\begin{document}

   \title{\hess\ detection of very-high-energy \gray\ emission \\from the quasar \pks}
   \titlerunning{\hess\ observations of \pks}

\author{H.E.S.S. Collaboration
\and H.~Abdalla \inst{\ref{NWU}}
\and R.~Adam \inst{\ref{LLR}}
\and F.~Aharonian \inst{\ref{MPIK},\ref{DIAS},\ref{RAU}}
\and F.~Ait~Benkhali \inst{\ref{MPIK}}
\and E.O.~Ang\"uner \inst{\ref{CPPM}}
\and M.~Arakawa \inst{\ref{Rikkyo}}
\and C.~Arcaro \inst{\ref{NWU}}
\and C.~Armand \inst{\ref{LAPP}}
\and H.~Ashkar \inst{\ref{IRFU}}
\and M.~Backes \inst{\ref{UNAM},\ref{NWU}}
\and V.~Barbosa~Martins \inst{\ref{DESY}}
\and M.~Barnard \inst{\ref{NWU}}
\and Y.~Becherini \inst{\ref{Linnaeus}}
\and D.~Berge \inst{\ref{DESY}}
\and K.~Bernl\"ohr \inst{\ref{MPIK}}
\and R.~Blackwell \inst{\ref{Adelaide}}
\and M.~B\"ottcher \inst{\ref{NWU}}
\and C.~Boisson \inst{\ref{LUTH}}
\and J.~Bolmont \inst{\ref{LPNHE}}
\and S.~Bonnefoy \inst{\ref{DESY}}
\and J.~Bregeon \inst{\ref{LUPM}}
\and M.~Breuhaus \inst{\ref{MPIK}}
\and F.~Brun \inst{\ref{IRFU}}
\and P.~Brun \inst{\ref{IRFU}}
\and M.~Bryan \inst{\ref{GRAPPA}}
\and M.~B\"{u}chele \inst{\ref{ECAP}}
\and T.~Bulik \inst{\ref{UWarsaw}}
\and T.~Bylund \inst{\ref{Linnaeus}}
\and M.~Capasso \inst{\ref{IAAT}}
\and S.~Caroff \inst{\ref{LPNHE}}
\and A.~Carosi \inst{\ref{LAPP}}
\and S.~Casanova \inst{\ref{IFJPAN},\ref{MPIK}}
\and M.~Cerruti\protect\footnotemark[1] \inst{\ref{LPNHE}, \ref{CerrutiNowAt}}
\and T.~Chand \inst{\ref{NWU}}
\and S.~Chandra \inst{\ref{NWU}}
\and A.~Chen \inst{\ref{WITS}}
\and S.~Colafrancesco${}^\dagger$  \inst{\ref{WITS}} 
\and M.~Cury{\l}o \inst{\ref{UJK}}
\and I.D.~Davids \inst{\ref{UNAM}}
\and C.~Deil \inst{\ref{MPIK}}
\and J.~Devin \inst{\ref{CENBG}}
\and P.~deWilt \inst{\ref{Adelaide}}
\and L.~Dirson \inst{\ref{HH}}
\and A.~Djannati-Ata\"i \inst{\ref{APC}}
\and A.~Dmytriiev \inst{\ref{LUTH}}
\and A.~Donath \inst{\ref{MPIK}}
\and V.~Doroshenko \inst{\ref{IAAT}}
\and L.O'C.~Drury \inst{\ref{DIAS}}
\and J.~Dyks \inst{\ref{NCAC}}
\and K.~Egberts \inst{\ref{UP}}
\and G.~Emery \inst{\ref{LPNHE}}
\and J.-P.~Ernenwein \inst{\ref{CPPM}}
\and S.~Eschbach \inst{\ref{ECAP}}
\and K.~Feijen \inst{\ref{Adelaide}}
\and S.~Fegan \inst{\ref{LLR}}
\and A.~Fiasson \inst{\ref{LAPP}}
\and G.~Fontaine \inst{\ref{LLR}}
\and S.~Funk \inst{\ref{ECAP}}
\and M.~F\"u{\ss}ling \inst{\ref{DESY}}
\and S.~Gabici \inst{\ref{APC}}
\and Y.A.~Gallant \inst{\ref{LUPM}}
\and F.~Gat{\'e} \inst{\ref{LAPP}}
\and G.~Giavitto \inst{\ref{DESY}}
\and D.~Glawion \inst{\ref{LSW}}
\and J.F.~Glicenstein \inst{\ref{IRFU}}
\and D.~Gottschall \inst{\ref{IAAT}}
\and M.-H.~Grondin \inst{\ref{CENBG}}
\and J.~Hahn \inst{\ref{MPIK}}
\and M.~Haupt \inst{\ref{DESY}}
\and G.~Heinzelmann \inst{\ref{HH}}
\and G.~Henri \inst{\ref{Grenoble}}
\and G.~Hermann \inst{\ref{MPIK}}
\and J.A.~Hinton \inst{\ref{MPIK}}
\and W.~Hofmann \inst{\ref{MPIK}}
\and C.~Hoischen \inst{\ref{UP}}
\and T.~L.~Holch \inst{\ref{HUB}}
\and M.~Holler \inst{\ref{LFUI}}
\and D.~Horns \inst{\ref{HH}}
\and D.~Huber \inst{\ref{LFUI}}
\and H.~Iwasaki \inst{\ref{Rikkyo}}
\and M.~Jamrozy \inst{\ref{UJK}}
\and D.~Jankowsky \inst{\ref{ECAP}}
\and F.~Jankowsky \inst{\ref{LSW}}
\and A.~Jardin-Blicq\inst{\ref{MPIK}}
\and I.~Jung-Richardt \inst{\ref{ECAP}}
\and M.A.~Kastendieck \inst{\ref{HH}}
\and K.~Katarzy{\'n}ski \inst{\ref{NCUT}}
\and M.~Katsuragawa \inst{\ref{KAVLI}}
\and U.~Katz \inst{\ref{ECAP}}
\and D.~Khangulyan \inst{\ref{Rikkyo}}
\and B.~Kh\'elifi \inst{\ref{APC}}
\and J.~King \inst{\ref{LSW}}
\and S.~Klepser \inst{\ref{DESY}}
\and W.~Klu\'{z}niak \inst{\ref{NCAC}}
\and Nu.~Komin \inst{\ref{WITS}}
\and K.~Kosack \inst{\ref{IRFU}}
\and D.~Kostunin \inst{\ref{DESY}} 
\and M.~Kraus \inst{\ref{ECAP}}
\and G.~Lamanna \inst{\ref{LAPP}}
\and J.~Lau \inst{\ref{Adelaide}}
\and A.~Lemi\`ere \inst{\ref{APC}}
\and M.~Lemoine-Goumard \inst{\ref{CENBG}}
\and J.-P.~Lenain\protect\footnotemark[1] \inst{\ref{LPNHE}}
\and E.~Leser \inst{\ref{UP},\ref{DESY}}
\and C.~Levy \inst{\ref{LPNHE}}
\and T.~Lohse \inst{\ref{HUB}}
\and I.~Lypova \inst{\ref{DESY}}
\and J.~Mackey \inst{\ref{DIAS}}
\and J.~Majumdar \inst{\ref{DESY}}
\and D.~Malyshev \inst{\ref{IAAT}}
\and V.~Marandon \inst{\ref{MPIK}}
\and A.~Marcowith \inst{\ref{LUPM}}
\and A.~Mares \inst{\ref{CENBG}}
\and C.~Mariaud \inst{\ref{LLR}}
\and G.~Mart\'i-Devesa \inst{\ref{LFUI}}
\and R.~Marx \inst{\ref{MPIK}}
\and G.~Maurin \inst{\ref{LAPP}}
\and P.J.~Meintjes \inst{\ref{UFS}}
\and A.M.W.~Mitchell \inst{\ref{MPIK},\ref{MitchellNowAt}}
\and R.~Moderski \inst{\ref{NCAC}}
\and M.~Mohamed \inst{\ref{LSW}}
\and L.~Mohrmann \inst{\ref{ECAP}}
\and J.~Muller \inst{\ref{LLR}}
\and C.~Moore \inst{\ref{Leicester}}
\and E.~Moulin \inst{\ref{IRFU}}
\and T.~Murach \inst{\ref{DESY}}
\and S.~Nakashima  \inst{\ref{RIKKEN}}
\and M.~de~Naurois \inst{\ref{LLR}}
\and H.~Ndiyavala  \inst{\ref{NWU}}
\and F.~Niederwanger \inst{\ref{LFUI}}
\and J.~Niemiec \inst{\ref{IFJPAN}}
\and L.~Oakes \inst{\ref{HUB}}
\and P.~O'Brien \inst{\ref{Leicester}}
\and H.~Odaka \inst{\ref{Tokyo}}
\and S.~Ohm \inst{\ref{DESY}}
\and E.~de~Oña~Wilhelmi \inst{\ref{DESY}}
\and M.~Ostrowski \inst{\ref{UJK}}
\and I.~Oya \inst{\ref{DESY}}
\and M.~Panter \inst{\ref{MPIK}}
\and R.D.~Parsons \inst{\ref{MPIK}}
\and C.~Perennes \inst{\ref{LPNHE}}
\and P.-O.~Petrucci \inst{\ref{Grenoble}}
\and B.~Peyaud \inst{\ref{IRFU}}
\and Q.~Piel \inst{\ref{LAPP}}
\and S.~Pita \inst{\ref{APC}}
\and V.~Poireau \inst{\ref{LAPP}}
\and A.~Priyana~Noel \inst{\ref{UJK}}
\and D.A.~Prokhorov \inst{\ref{WITS}}
\and H.~Prokoph\protect\footnotemark[1] \inst{\ref{DESY}}
\and G.~P\"uhlhofer \inst{\ref{IAAT}}
\and M.~Punch \inst{\ref{APC},\ref{Linnaeus}}
\and A.~Quirrenbach \inst{\ref{LSW}}
\and S.~Raab \inst{\ref{ECAP}}
\and R.~Rauth \inst{\ref{LFUI}}
\and A.~Reimer \inst{\ref{LFUI}}
\and O.~Reimer \inst{\ref{LFUI}}
\and Q.~Remy \inst{\ref{LUPM}}
\and M.~Renaud \inst{\ref{LUPM}}
\and F.~Rieger \inst{\ref{MPIK}}
\and L.~Rinchiuso \inst{\ref{IRFU}}
\and C.~Romoli \inst{\ref{MPIK}}
\and G.~Rowell \inst{\ref{Adelaide}}
\and B.~Rudak \inst{\ref{NCAC}}
\and E.~Ruiz-Velasco \inst{\ref{MPIK}}
\and V.~Sahakian \inst{\ref{YPI}}
\and S.~Saito \inst{\ref{Rikkyo}}
\and D.A.~Sanchez \inst{\ref{LAPP}}
\and A.~Santangelo \inst{\ref{IAAT}}
\and M.~Sasaki \inst{\ref{ECAP}}
\and R.~Schlickeiser \inst{\ref{RUB}}
\and F.~Sch\"ussler \inst{\ref{IRFU}}
\and A.~Schulz \inst{\ref{DESY}}
\and H.~Schutte \inst{\ref{NWU}}
\and U.~Schwanke \inst{\ref{HUB}}
\and S.~Schwemmer \inst{\ref{LSW}}
\and M.~Seglar-Arroyo \inst{\ref{IRFU}}
\and M.~Senniappan \inst{\ref{Linnaeus}}
\and A.S.~Seyffert \inst{\ref{NWU}}
\and N.~Shafi \inst{\ref{WITS}}
\and K.~Shiningayamwe \inst{\ref{UNAM}}
\and R.~Simoni \inst{\ref{GRAPPA}}
\and A.~Sinha \inst{\ref{APC}}
\and H.~Sol \inst{\ref{LUTH}}
\and A.~Specovius \inst{\ref{ECAP}}
\and M.~Spir-Jacob \inst{\ref{APC}}
\and {\L.}~Stawarz \inst{\ref{UJK}}
\and R.~Steenkamp \inst{\ref{UNAM}}
\and C.~Stegmann \inst{\ref{UP},\ref{DESY}}
\and C.~Steppa \inst{\ref{UP}}
\and T.~Takahashi  \inst{\ref{KAVLI}}
\and T.~Tavernier \inst{\ref{IRFU}}
\and A.M.~Taylor \inst{\ref{DESY}}
\and R.~Terrier \inst{\ref{APC}}
\and D.~Tiziani \inst{\ref{ECAP}}
\and M.~Tluczykont \inst{\ref{HH}}
\and C.~Trichard \inst{\ref{LLR}}
\and M.~Tsirou \inst{\ref{LUPM}}
\and N.~Tsuji \inst{\ref{Rikkyo}}
\and R.~Tuffs \inst{\ref{MPIK}}
\and Y.~Uchiyama \inst{\ref{Rikkyo}}
\and D.J.~van~der~Walt \inst{\ref{NWU}}
\and C.~van~Eldik \inst{\ref{ECAP}}
\and C.~van~Rensburg \inst{\ref{NWU}}
\and B.~van~Soelen \inst{\ref{UFS}}
\and G.~Vasileiadis \inst{\ref{LUPM}}
\and J.~Veh \inst{\ref{ECAP}}
\and C.~Venter \inst{\ref{NWU}}
\and P.~Vincent \inst{\ref{LPNHE}}
\and J.~Vink \inst{\ref{GRAPPA}}
\and F.~Voisin \inst{\ref{Adelaide}}
\and H.J.~V\"olk \inst{\ref{MPIK}}
\and T.~Vuillaume \inst{\ref{LAPP}}
\and Z.~Wadiasingh \inst{\ref{NWU}}
\and S.J.~Wagner \inst{\ref{LSW}}
\and R.~White \inst{\ref{MPIK}}
\and A.~Wierzcholska \inst{\ref{IFJPAN},\ref{LSW}}
\and R.~Yang \inst{\ref{MPIK}}
\and H.~Yoneda \inst{\ref{KAVLI}}
\and M.~Zacharias \inst{\ref{NWU}}
\and R.~Zanin \inst{\ref{MPIK}}
\and A.A.~Zdziarski \inst{\ref{NCAC}}
\and A.~Zech \inst{\ref{LUTH}}
\and A.~Ziegler \inst{\ref{ECAP}}
\and J.~Zorn \inst{\ref{MPIK}}
\and N.~\.Zywucka \inst{\ref{NWU}}
\and P. S.~Smith \inst{\ref{Arizona}}
}

\institute{
Centre for Space Research, North-West University, Potchefstroom 2520, South Africa \label{NWU} \and 
Laboratoire Leprince-Ringuet, Ecole Polytechnique, CNRS/IN2P3, F-91128 Palaiseau, France \label{LLR} \and
Max-Planck-Institut f\"ur Kernphysik, P.O. Box 103980, D 69029 Heidelberg, Germany \label{MPIK} \and 
Dublin Institute for Advanced Studies, 31 Fitzwilliam Place, Dublin 2, Ireland \label{DIAS} \and 
High Energy Astrophysics Laboratory, RAU,  123 Hovsep Emin St  Yerevan 0051, Armenia \label{RAU} \and
Aix Marseille Universit\'e, CNRS/IN2P3, CPPM, Marseille, France \label{CPPM} \and
Department of Physics, Rikkyo University, 3-34-1 Nishi-Ikebukuro, Toshima-ku, Tokyo 171-8501, Japan \label{Rikkyo} \and
Laboratoire d'Annecy de Physique des Particules, Univ. Grenoble Alpes, Univ. Savoie Mont Blanc, CNRS, LAPP, 74000 Annecy, France \label{LAPP} \and
IRFU, CEA, Universit\'e Paris-Saclay, F-91191 Gif-sur-Yvette, France \label{IRFU} \and
University of Namibia, Department of Physics, Private Bag 13301, Windhoek, Namibia, 12010 \label{UNAM} \and
DESY, D-15738 Zeuthen, Germany \label{DESY} \and
Department of Physics and Electrical Engineering, Linnaeus University,  351 95 V\"axj\"o, Sweden \label{Linnaeus} \and
School of Physical Sciences, University of Adelaide, Adelaide 5005, Australia \label{Adelaide} \and
LUTH, Observatoire de Paris, PSL Research University, CNRS, Universit\'e Paris Diderot, 5 Place Jules Janssen, 92190 Meudon, France \label{LUTH} \and
Sorbonne Universit\'e, Universit\'e Paris Diderot, Sorbonne Paris Cit\'e, CNRS/IN2P3, Laboratoire de Physique Nucl\'eaire et de Hautes Energies, LPNHE, 4 Place Jussieu, F-75252 Paris, France \label{LPNHE} \and
Laboratoire Univers et Particules de Montpellier, Universit\'e Montpellier, CNRS/IN2P3,  CC 72, Place Eug\`ene Bataillon, F-34095 Montpellier Cedex 5, France \label{LUPM} \and
GRAPPA, Anton Pannekoek Institute for Astronomy, University of Amsterdam,  Science Park 904, 1098 XH Amsterdam, The Netherlands \label{GRAPPA} \and
Friedrich-Alexander-Universit\"at Erlangen-N\"urnberg, Erlangen Centre for Astroparticle Physics, Erwin-Rommel-Str. 1, D 91058 Erlangen, Germany \label{ECAP} \and
Astronomical Observatory, The University of Warsaw, Al. Ujazdowskie 4, 00-478 Warsaw, Poland \label{UWarsaw} \and
Institut f\"ur Astronomie und Astrophysik, Universit\"at T\"ubingen, Sand 1, D 72076 T\"ubingen, Germany \label{IAAT} \and
Instytut Fizyki J\c{a}drowej PAN, ul. Radzikowskiego 152, 31-342 Krak{\'o}w, Poland \label{IFJPAN} \and
School of Physics, University of the Witwatersrand, 1 Jan Smuts Avenue, Braamfontein, Johannesburg, 2050 South Africa \label{WITS} \and
Obserwatorium Astronomiczne, Uniwersytet Jagiello{\'n}ski, ul. Orla 171, 30-244 Krak{\'o}w, Poland \label{UJK} \and 
Universit\'e Bordeaux, CNRS/IN2P3, Centre d'\'Etudes Nucl\'eaires de Bordeaux Gradignan, 33175 Gradignan, France \label{CENBG} \and
Universit\"at Hamburg, Institut f\"ur Experimentalphysik, Luruper Chaussee 149, D 22761 Hamburg, Germany \label{HH} \and
APC, AstroParticule et Cosmologie, Universit\'{e} Paris Diderot, CNRS/IN2P3, CEA/Irfu, Observatoire de Paris, Sorbonne Paris Cit\'{e}, 10, rue Alice Domon et L\'{e}onie Duquet, 75205 Paris Cedex 13, France \label{APC} \and
Nicolaus Copernicus Astronomical Center, Polish Academy of Sciences, ul. Bartycka 18, 00-716 Warsaw, Poland \label{NCAC} \and
Institut f\"ur Physik und Astronomie, Universit\"at Potsdam,  Karl-Liebknecht-Strasse 24/25, D 14476 Potsdam, Germany \label{UP} \and
Landessternwarte, Universit\"at Heidelberg, K\"onigstuhl, D 69117 Heidelberg, Germany \label{LSW} \and
Univ. Grenoble Alpes, CNRS, IPAG, F-38000 Grenoble, France \label{Grenoble} \and
Institut f\"ur Physik, Humboldt-Universit\"at zu Berlin, Newtonstr. 15, D 12489 Berlin, Germany \label{HUB} \and
Institut f\"ur Astro- und Teilchenphysik, Leopold-Franzens-Universit\"at Innsbruck, A-6020 Innsbruck, Austria \label{LFUI} \and
Centre for Astronomy, Faculty of Physics, Astronomy and Informatics, Nicolaus Copernicus University,  Grudziadzka 5, 87-100 Torun, Poland \label{NCUT} \and
Kavli Institute for the Physics and Mathematics of the Universe (Kavli IPMU), The University of Tokyo Institutes for Advanced Study (UTIAS), The University of Tokyo, 5-1-5 Kashiwa-no-Ha, Kashiwa City, Chiba, 277-8583, Japan \label{KAVLI} \and
Department of Physics, University of the Free State,  PO Box 339, Bloemfontein 9300, South Africa \label{UFS} \and
Department of Physics and Astronomy, The University of Leicester, University Road, Leicester, LE1 7RH, United Kingdom \label{Leicester} \and
RIKEN, 2-1 Hirosawa, Wako, Saitama 351-0198, Japan \label{RIKKEN} \and
Department of Physics, The University of Tokyo, 7-3-1 Hongo, Bunkyo-ku, Tokyo 113-0033, Japan \label{Tokyo} \and
Yerevan Physics Institute, 2 Alikhanian Brothers St., 375036 Yerevan, Armenia \label{YPI} \and
Institut f\"ur Theoretische Physik, Lehrstuhl IV: Weltraum und Astrophysik, Ruhr-Universit\"at Bochum, D 44780 Bochum, Germany \label{RUB} \and
Steward Observatory, University of Arizona, Tucson, AZ 85721, USA \label{Arizona} \and
Now at Institut de Ci\`{e}ncies del Cosmos (ICC UB), Universitat de Barcelona (IEEC-UB), Mart\'{i} Franqu\`es 1, E08028 Barcelona, Spain\label{CerrutiNowAt} \and 
Now at Physik Institut, Universit\"at Z\"urich, Winterthurerstrasse 190, CH-8057 Z\"urich, Switzerland \label{MitchellNowAt} 
}

\offprints{H.E.S.S.~collaboration,
\protect\\\email{\href{mailto:contact.hess@hess-experiment.eu}{contact.hess@hess-experiment.eu}};
\protect\\\protect\footnotemark[1] Corresponding authors
\protect\\${}^\dagger$ Deceased
}

    \date{\today}
 \abstract{
\tiny Flat-spectrum radio-quasars (FSRQs) are rarely detected at very-high-energies (VHE; E $\geq$ 100 GeV) due to their low-frequency-peaked spectral energy distributions. At present, only six FSRQs are known to emit VHE photons, representing only 7$\%$ of the VHE extragalactic catalog, which is largely dominated by high-frequency-peaked BL Lacertae objects.}{Following the detection of MeV-GeV \gray\ flaring activity from the FSRQ \pks\ (z=0.189) with \fermilat, the H.E.S.S. array of Cherenkov telescopes triggered target-of-opportunity (ToO) observations on February 18, 2015, with the goal of studying the \gray\ emission in the VHE band.}{\hess\ ToO observations were carried out during the nights of February 18, 19, 21, and 24, 2015. Together with \fermilat\, the multi-wavelength coverage of the flare includes \swift\ observations in soft-X-rays and optical/UV, and optical monitoring (photometry and spectro-polarimetry) by the Steward Observatory, the ATOM, the KAIT and the ASAS-SN telescope.}{VHE emission from \pks\ was detected with \hess\ during the night of February 19, 2015, only. \fermilat\ data indicate the presence of a \gray\ flare, peaking at the time of the \hess\ detection, with a flux doubling time-scale of around six hours. The \gray\ flare was accompanied by at least a 1 mag brightening of the
non-thermal optical continuum. No simultaneous observations at longer wavelengths are available for the night of the \hess\ detection. The \gray\ observations with \hess\ and \fermilat\ are used to put constraints on the location of the \gray\ emitting region during the flare: it is constrained to be just outside the radius of the broad-line-region $r_{BLR}$ with a bulk Lorentz factor $\Gamma \simeq 20$, or at the level of the radius of the dusty torus $r_{torus}$ with $\Gamma \simeq 60$.}{\pks\ is the seventh FSRQ known to emit VHE photons and, at z=0.189, is the nearest so far. The location of the \gray\ emitting region during the flare can be tightly constrained thanks to opacity, variability, and collimation arguments.}


   \maketitle

\makeatletter
\renewcommand*{\@fnsymbol}[1]{\ifcase#1\@arabic{#1}\fi}
\makeatother

%
  \def\linenumberfont{\normalfont\scriptsize}


\section{Introduction}
\label{intro}

The very-high-energy (VHE; E $\geq$ 100 GeV) window on the Universe was opened with the discovery of VHE emission from the Crab Nebula \citep{Weekes89} using the \textit{Whipple} 10-m Imaging Atmospheric Cherenkov Telescope (IACT). Soon after, the first extragalactic VHE source, the blazar Markarian 421, was discovered by \citet{Punch92}. A few decades later, thanks to the current generation of IACTs (H.E.S.S., MAGIC and VERITAS), the number of known VHE extragalactic sources has grown to 82\footnote{In October 2019. For an up-to-date list of VHE sources see \url{http://tevcat.uchicago.edu}, \citet{tevcat}}.  The majority of them (around 90\%) are blazars.\\

Within the current unification model of active galactic nuclei (AGNs), blazars are interpreted as radio-loud AGNs whose relativistic jet points in the direction of the observer \citep[see][]{Blandford78}. From an observational point of view, two sub-classes of blazars exist: flat-spectrum radio-quasars (FSRQs) and BL Lacertae objects, according to the equivalent width of emission lines from the Broad-Line Region (BLR), which is $> 5\ \AA$ in FSRQs \citep[see e.g.][]{Urry95}. All blazars are characterized by a similar spectral energy distribution (SED), which consists of two distinct components peaking in the infrared-to-X-ray band and the MeV-to-TeV band, respectively \citep[see e.g.][]{FermiSED}. While FSRQs are in general characterized by a relatively low frequency (in the infrared) of the low-energy SED peak, BL Lac objects are further classified by the peak frequency of their first SED component into low / intermediate / high frequency-peaked BL Lac objects \citep[LBLs / IBLs / HBLs,][]{Padovani95, Sambruna96}. Hence, observations in a narrow frequency window unavoidably preselect a particular blazar sub-class, and it is not a surprise that observations at VHE \graynohy s detect more likely HBLs whose overall spectrum can peak in the VHE band at energies up to few TeVs  \citep[such as the extreme HBL 1ES~0229+200, see][]{0229hess, 0229veritas}, and represent about 70$\%$ of extragalactic VHE sources. In addition, FSRQs and BL Lac objects have different redshift distributions, with the former being located on average at larger distances \citep[see][]{Padovani92}. VHE astronomy has the important property of being limited in redshift, due to the absorption of VHE photons via pair-production over the extragalactic background light \citep[EBL, see][]{Salamon98}. Thus, both intrinsic source properties and propagation effects make FSRQs difficult to be observed with IACTs. 
So far only six FSRQs have been detected by IACTs: 3C~279 \citep{3C279magic, HESS3C279}, PKS~1222+216 \citep{PKS1222magic, cerruti1222}, PKS~1510-089 \citep{1510hess, 1510magic}, PKS~1441+252 \citep{1441veritas, 1441magic}, S3~0218+35 \citep{0218magic}, which is currently the most distant source of VHE photons ever observed (z=0.944), and Ton~599 \citep{Ton599magic, Ton599veritas}. With the notable exception of PKS~1510-089 \citep{Magic1510lowstate}, which at z=0.361 is the nearest among them, all other FSRQs have been detected at VHE only during bright flaring activity, and are all characterized by very soft VHE spectra.\\

The radiation mechanism responsible for the low-energy SED component of blazars is thought to be synchrotron emission by a non-thermal population of leptons (electrons and positrons) in the jet. 
The radiation mechanism responsible for the \gray\ emission is thought to be inverse-Compton scattering off low-energy photons by the same leptons which produce the synchrotron SED component. For FSRQs, the low-energy target photons are thought to be thermal photons from the accretion disc, or from the dusty torus, or emission lines produced in the BLR \citep[see][]{dermer93, sikora94, blaz00}. This type of scenario, called External-Inverse-Compton (EIC), proved to be able to successfully reproduce the broad-band SED of \gray\ FSRQs \citep[see e.g.][]{ghisellini10, meyer12, bottcher13}. Alternative hadronic emission scenarios, although capable of modeling the SED of FSRQs \citep[see e.g.][]{bottcher13},  encounter difficulties due to the high power required to reproduce the photon emission \citep{sikora09, reimer12, petro15, zdz15}.\\

Even though there is general consensus on the EIC as the emission mechanism, several questions remain open. Among them is the uncertainty as to where the \gray\ emission region is located within the relativistic jet. Does the emission come from the base of the jet, near the supermassive black hole (SMBH) powering the quasar, or is it rather produced downstream in the jet? The answer to this question, which arose with the detection of the first \gray\ AGNs by EGRET \citep[see e.g.,][]{dermer92, dermer94, becker95, blandford95, marcowith95, jorstad01},  is still a major active research topic in blazar physics and has been adressed by several authors in the last years \citep[see e.g.,][]{ghisellini09, finke10, poutanen10, tavecchio10, agudo11a, agudo11b, hayashida12, yan12, caowang13, rani13, brown13, dermer14, maxm14, nalew14, dotson15, ramak15, coogan16, finke16}. In this paper it is shown that, making use only of \gray\ observations of the quasar \pks, it is possible to put a tight constraint on the location of the \gray\ emitting region within the jet.\\

The quasar \pks\ was first detected as a radio-source with the Parkes telescope \citep{Day66}. Its radio morphology is typical of a blazar, with a compact core and a single-sided, parsec-scale jet \citep{Lister05, Lister09}. The optical-UV spectrum is characterized by broad emission lines and a big-blue bump, associated with thermal emission from the SMBH accretion disk \citep{Baldwin75, Malkan86}. Thanks to the emission lines, the redshift of \pks\ is well determined: the most recent measurement is z = 0.189 \citep{Ho09}. As is typical for blazars, a giant elliptical galaxy hosts the AGN \citep{Kotilainen98, Wright98, Mclure99}. During January 2002, \pks\ exhibited an extremely bright and fast optical flare \citep[0.6 magnitudes/hour, see][]{Clements03}, which classifies the source also as an Optically Violently Variable (OVV) quasar.\\

In high-energy \graynohy s (HE; 100 MeV $\leq$ E $\leq$ 100 GeV),  \pks\ is detected by \fermilat, and is included in the most recent \fermilat\ catalog \citep[3FGL,][]{3FGL} under the name \textit{3FGL~J0739.4+0137}. Since the beginning of the \textit{Fermi} mission in 2008, the source remained relatively quiescent until November 2014, when a \gray\ flare was detected \citep{FermiAtelNov}. The source remained active in HE \graynohy s for the following months, and reached the maximum \gray\ flux in February 2015. The flaring state in HE \graynohy s triggered VHE observations with the High Energy Stereoscopic System (\hess), resulting in the first detection of VHE photons from \pks, which is reported in this paper.\\

The paper is organized as follows: in Section \ref{section2} are presented the observations of \pks\ and the data analysis, for \hess\ and for lower-energy instruments; in Section \ref{section3}  are discussed the implications of the VHE detection, in particular in terms of the location of the \gray\ emitting region, and  is presented the SED of the source during the flare; conclusions are in Section \ref{section4}.\\

\section{Observations and Data Analysis}
\label{section2}

\subsection{H.E.S.S}
\hess\ is an array of IACTs located in the Khomas Highland of Namibia (23$^\circ$16\arcmin 18\arcsec~S, 16$^\circ$30\arcmin 00\arcsec E), at an altitude of about 1800~m above sea level. As all other IACTs, \hess\ images the Cherenkov light emitted by particle showers triggered by interaction of \graynohy s with the Earth's atmosphere. The study of the shower images makes it possible to reconstruct the incoming direction of a \graynohy , its energy and its arrival time. 
The array consists of four 12-m diameter reflectors arranged in a square of 120~m side length and, since 2012, one additional 28-m diameter telescope (in the following called CT5) in the middle of the array \citep{hessiiagn}. This hybrid configuration of the array allows data to be taken in different modes by triggering on events either detected by CT5 only (monoscopic mode) or by any combination of two or more telescopes (stereoscopic mode). The standard observation mode is to collect both monoscopic and stereoscopic events during the same observation to allow for a low-energy analysis threshold (below 100~GeV in the monoscopic mode) as well as a good spatial and spectral reconstruction based on the excellent background rejection power in the stereoscopic mode. \\

\begin{figure}[t!]
\begin{center}
\includegraphics[width=270pt]{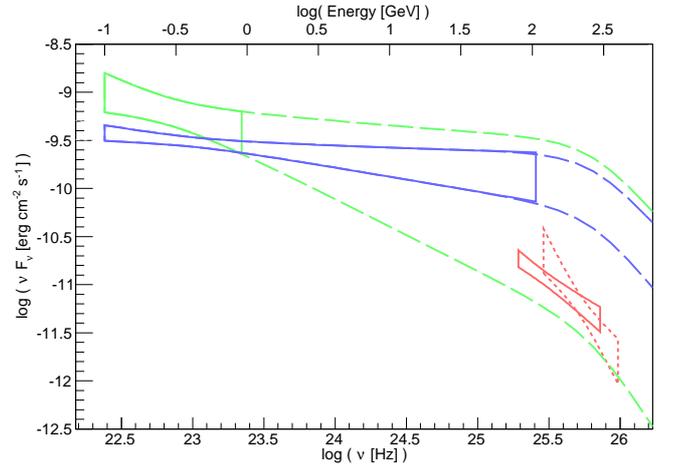}
\caption{Gamma-ray emission from \pks\ centred on the night of February 19, 2015. The red bow-tie represents the spectrum measured with \hess\, using the monoscopic (bold line) and the stereoscopic analysis (dashed line). The green and blue bow-ties represent the spectra measured with \fermilat\ strictly simultaneous with the \hess\ detection, and integrating over an exposure of 24 hours (MJD 57072.5-57073.5) around the \hess\ detection, respectively; the extrapolation of \fermilat\ spectra to higher energies (green and blue long-dashed lines) takes into account the absorption on the EBL \citep{franceschini08}.\label{gammaspectra}}
\end{center}
\end{figure}

\begin{table*}[t!]
   \begin{center}
   \caption{Details of the \hess\ observations of \pks.  \label{TableHESS}
}
\begin{tabular}{|c|c|c|c|}
\hline
Array configuration & Starting time & Exposure & Flux (E $>100$ GeV) \\
 & [UTC] & [h]  & [$10^{-11}$ cm$^{-2}$ s$^{-1}$] \\
 \hline
CT1-5 & Feb 18, 2015, 21:21 & 1.8 & $< 3.6$ \\
\hline
CT1-5 & Feb 19, 2015, 18:53 & 1.8 & $5.7 \pm 1.1$  \\
\hline
CT1-5 & Feb 21, 2015, 20:31 & 2.7 & $< 2.0$\\
\hline
CT5    & Feb 24, 2015, 21:43 & 0.9 &  $< 3.1$ \\
\hline
\end{tabular}
\end{center}
\end{table*}

Target of Opportunity (ToO) observations of \pks\ with \hess\ were triggered on February 18, 2015, following the detection of flaring activity in \fermilat\ data (see next Section). \hess\ observations were carried out with the full array in the same night. Follow-up observations were performed on February 19, 21, and 24 (with the full array for the first two nights, while data were recorded with CT5 only during the last night). All observations were taken in ``wobble mode'' where the source position is offset by 0\fdg{5} from the camera center to allow for simultaneous background estimation \citep{berge07}. The details of \hess\ obsevations are reported in Table \ref{TableHESS}.\\

Data were analyzed with the ImPACT analysis chain \citep{2014APh....56...26P,2015ICRC...34..826P} using both monoscopic and stereoscopic reconstruction and were crosschecked with an independent analysis chain \citep{2009APh....32..231D,2015arXiv150902896H} yielding consistent results. With the \textit{loose configuration} of the monoscopic analysis, an excess of 473 $\gamma$-like events with a signal-to-background ratio of 0.17 is found at the source position in the overall good-quality 7.2~hour dataset, corresponding to a detection significance of 8.5$\sigma$ using Eq.~17 of \citet{1983ApJ...272..317L}. The overall signal is dominated by the strong excess coming from the second night of observations (February 19, 2015). The monoscopic (respectively, stereoscopic) analysis of this 1.8~h live time data set yields an excess of 364 (49) $\gamma$-like events, corresponding to an 11.1$\sigma$ (5.3$\sigma$) post-trial\footnote{Four trials are considered for the computation of significances: two for the analyses (monoscopic and stereoscopic), and two for the time selection (the full data-set, and the night of February 19, 2015).} significance, while the source is not detected with a significance greater than 5$\sigma$ in any other night.\\ 

To determine the position of the VHE \gray\ emission, a two-dimensional fit to the \gray\ excess (for the night of the VHE detection only) is performed using the stereoscopic data set, yielding a shape consistent with a point-like source. The best-fit position is found at RA(J2000) = 07h 39m 17s $\pm$ 3s, DEC(J2000) = 01$^\circ$ 36\arcmin 29\arcsec $\pm$ 56\arcsec, and is positionally consistent with the radio position of  \pks\ \citep{2010AJ....139.1695L}.\\ 

The differential energy spectrum of the \gray\ emission is derived by performing a spectral fit, again for the night of the detection only. Both monoscopic and stereoscopic spectra are consistent with a power-law model of the form $dN/dE = N_0 (E/E_0)^{-\Gamma}$. The photon index is estimated to be $\Gamma = 3.1\pm0.3_\mathrm{stat}\pm 0.2_\mathrm{syst}$ for the monoscopic analysis and $\Gamma = 4.2\pm0.8_\mathrm{stat}\pm 0.2_\mathrm{syst}$ for the stereoscopic one, and the flux normalization at 200~GeV is found to be $N_0 = (1.0\pm0.2_\mathrm{stat}\pm 0.3_\mathrm{syst}) \times 10^{-10}\ \mathrm{cm}^{-2}\ \mathrm{s}^{-1}\ \mathrm{TeV}^{-1}$ and $N_0 = (1.1\pm0.3_\mathrm{stat}\pm 0.2_\mathrm{syst}) \times 10^{-10}\ \mathrm{cm}^{-2}\ \mathrm{s}^{-1}\ \mathrm{TeV}^{-1}$, respectively.
It is important to underline that the two analyses have different energy thresholds, equal to 80 and 150 GeV, respectively. The spectral results obtained are presented in Fig. \ref{gammaspectra}.\\

\begin{figure*}[t!]
\begin{center}
\begin{tikzpicture}
    \node[anchor=south west,inner sep=0] (image) at (0,0) {\includegraphics[width=290pt]{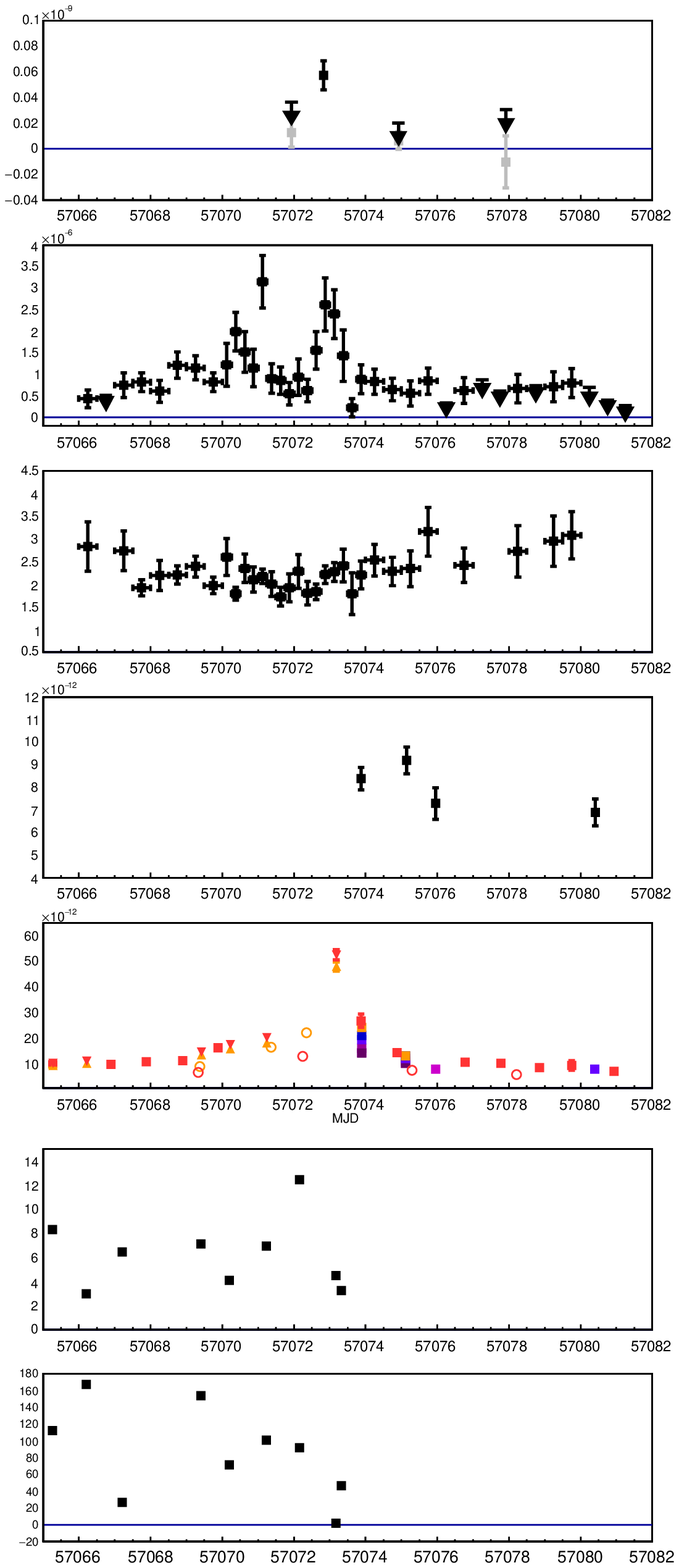}};
    \begin{scope}[x={(image.south east)},y={(image.north west)}]
                \draw[red,thick,dashed] (0.47,0.025) -- (0.47,0.985);
    \end{scope}
    \node[] at (8.9,20.2) {a};
    \node[] at (8.9,17.3) {b};
    \node[] at (8.9,14.3) {c};
    \node[] at (8.9,11.35) {d};
    \node[] at (8.9,8.4) {e};
    \node[] at (8.9,5.4) {f};
    \node[] at (8.9,2.45) {g};
    \node[scale=0.7] at (2.9,20.55) {15/02/15};
    \node[scale=0.7] at (4.8,20.55) {15/02/19};
    \node[scale=0.7] at (6.65, 20.55) {15/02/23};
    \node[scale=0.7, rotate=90] at (0.3,19.25) {HESS Flux (E$>$0.1 TeV)};
        \node[scale=0.7, rotate=90] at (0.55,19.25) {[cm$^{-2}$ s$^{-1}$]};
    \node[scale=0.7, rotate=90] at (0.3,16.3) {LAT Flux (E$>$0.1 GeV)};
        \node[scale=0.7, rotate=90] at (0.55,16.3) {[cm$^{-2}$ s$^{-1}$]};
    \node[scale=0.7, rotate=90] at (0.3,13.3) {LAT photon index};
    \node[scale=0.7, rotate=90] at (0.3,10.35) {XRT Flux (0.3-10 keV)};
        \node[scale=0.7, rotate=90] at (0.55,10.35) {[erg cm$^{-2}$ s$^{-1}$]};
    \node[scale=0.7, rotate=90] at (0.3,7.45) {Optical-UV Flux };
        \node[scale=0.7, rotate=90] at (0.55,7.45) {[erg cm$^{-2}$ s$^{-1}$]};
    \node[scale=0.7, rotate=90] at (0.3,4.45) {Polarization fraction [$\%$]};
    \node[scale=0.7, rotate=90] at (0.3,1.57) {Polarization angle [degrees]};
        \node[scale=0.7] at (5.05, 0.05) {MJD};
\end{tikzpicture}
\caption{Multi-wavelength light curve of \pks\ during February 2015. Panels from a to g: \hess\ integral flux above 100 GeV from the monoscopic analysis (in gray, the flux for the nights with no significant detection); \fermilat\ integral flux above 100 MeV; \fermilat\ photon index; \xrt\ integral flux between 0.3 and 10 keV, corrected for absorption; \uvot\ (orange, light-blue, blue, light-violet, violet and dark-violet squares for the V,B,U,W1,M2 and W2 filter, respectively), ATOM (red squares, for the R filter), Steward Observatory (orange and red triangles, for the V and R filter, respectively), ASAS-SN (orange open circles, for the V filter), and KAIT (red open circles, unfiltered) fluxes, de-reddened; Steward Observatory broad-band polarization fraction; and Steward Observatory polarization position angle (both averaged over 5000-7000 $\AA$). The vertical dashed red line indicates the time of the \hess\ detection.  \label{LCMWL}}
\end{center}
\end{figure*}

The night-by-night light curve for the monoscopic reconstruction, provided as an integral flux above 100 GeV, assuming a photon index of $3.1$, is shown in Fig.~\ref{LCMWL}, panel \lq\lq a\rq\rq. For the nights of no detection with \hess, upper limits on the VHE emission (estimated following \citet{rolke05} at the 95$\%$ confidence level) are also shown. No evidence for intra-night variability is found in the \hess\ data during the night of February 19, 2015: a fit of the 28-minutes-binned light curve with a constant function results in a $\chi^2$ value of 1.9 for three degrees of freedom. \\

\begin{table*}[t!]
   \begin{center}
   \caption{Details of the \swift\ observations of \pks.  \label{Tableswift}
}
\begin{tabular}{|c|c|c|c|c|c|c|}
\hline
Obs. ID & Starting time & Exposure & Power-law & Flux 0.3-10 keV & $\chi^2$/DOF & UVOT filters \\
 & [UTC] & [ks]  &  Index & [10$^{-12}$ erg cm$^{-2}$ s$^{-1}$]  & & \\
 \hline
00033535009 & Feb 20, 2015, 19:26 & 3.4 & 1.62 $\pm$ 0.08 & 8.4 $\pm$ 0.5 & 11.2 / 14 & V,B,U,W1,W2,M2 \\
\hline
00033535010 & Feb 22, 2015, 01:46 & 2.5 &  1.49 $\pm$ 0.09 & 9.2 $\pm$ 0.6 & 13.7 / 10 & V,B,U,W1,W2,M2 \\
\hline
00033535011 & Feb 22, 2015, 21:25 & 2.6 &  1.46 $\pm$ 0.13 & 7.3 $\pm$ 0.7 & 5.4 / 6 & W1 \\
\hline
00033535012 & Feb 27, 2015, 08:01 & 3.8 & 1.34 $\pm$ 0.09 & 6.9 $\pm$ 0.6 & 9.4 / 11 & U \\
\hline
\end{tabular}
\end{center}
\end{table*}

\subsection{\fermilat}
Detecting \gray\ photons with energies between 20\,MeV and above 300\,GeV, the LAT instrument \citep{2009ApJ...697.1071A} onboard the \textit{Fermi} satellite monitors the high-energy \gray\ sky every three hours. This instrument is thus ideal to reveal active states in AGNs, which could be used to trigger observations with other facilities as ToO observations. On February 18, 2015, such an active state was detected in \pks\ using the public monitored source list\footnote{\href{https://fermi.gsfc.nasa.gov/ssc/data/access/lat/msl_lc/}{https://fermi.gsfc.nasa.gov/ssc/data/access/lat/msl\_lc/}} as well as the dedicated FLaapLUC aperture-photometry pipeline \citep{flaapluc}. Following this early flare detection, a ToO campaign was launched with \hess, as reported in the previous Section.\\

The \fermilat\ data are analyzed with the public ScienceTools \texttt{v10r0p5}\footnote{See \href{http://fermi.gsfc.nasa.gov/ssc/data/analysis/documentation}{http://fermi.gsfc.nasa.gov/ssc/data/analysis/documentation}.}, and events are selected within a circular region of interest of 10\degr\ in radius centred on the nominal position of 3FGL\,J0739.4+0137, in order to perform a binned analysis as implemented in the \texttt{gtlike} tool. To encompass the entire active state studied here, data between February 1, 2015 and April 1, 2015 are considered, in the 100\,MeV--500\,GeV energy range. The \texttt{P8R3\_SOURCE\_V6} instrument response functions were used, together with a zenith angle cut of 90\degr\ to avoid contamination by the \gray\ bright Earth limb emission. The model of the region of interest was built based on the 3FGL catalog \citep{3FGL}, and it has been checked \textit{a posteriori} that no significant residual remains, which could have hinted for new sources not referenced in the 3FGL catalogue. The Galactic diffuse emission has been modeled using the file \texttt{gll\_iem\_v06.fits} \citep{2016ApJS..223...26A} and the isotropic background using \texttt{iso\_P8R3\_SOURCE\_V6\_v06.txt}.\\

For the time window considered here, the spectrum of
\pks\ is best described using a log-parabolic shape.\footnote{The best-fit \gray\ spectrum of 3FGL\,J0739.4+0137 as provided in the 3FGL catalog is a log-parabola with index $a=2.25\pm 0.05$, curvature $b=0.17\pm 0.03$, reference energy $E_0 = 327.1$ MeV and $E F_{(1\,\rm{GeV} - 100\,\rm{GeV})}=(3.66 \pm 0.13) \times 10^{-11}$\,erg\,cm$^{-2}$\,s$^{-1}$. The \fermilat\ Collaboration has recently released a new \gray\ catalog, the 4FGL \citep{4FGL}. The 4FGL associated source  for PKS 0736+017, 4FGL\,J0739.2+0137, is again best described by a log-parabolic spectrum, with $a=2.33\pm 0.02$, curvature $b=0.09\pm 0.01$, reference energy $E_0 = 503.3$ MeV  and $E F_{(100\,\rm{MeV} - 100\,\rm{GeV})}=(6.12 \pm 0.14) \times 10^{-11}$\,erg\,cm$^{-2}$\,s$^{-1}$.} A comparison with a power-law spectrum yielded a log-likelihood ratio of $27.5$ in favor of the log-parabola. Under this spectral hypothesis, \pks\ is detected with a test statistic \citep[TS;][]{1996ApJ...461..396M} of 2061, i.e. $\sim$45$\sigma$, with a flux $F_{(100\,\rm{MeV} - 500\,\rm{GeV})}=(5.42 \pm 0.27) \times 10^{-7}$\,cm$^{-2}$\,s$^{-1}$, index $a=2.22 \pm 0.06$, curvature index $b=0.037 \pm 0.025$, and reference energy $E_0 = 327$ MeV, averaged over the 2 months of
data considered here. In Fig.~\ref{LCMWL}, panel \lq\lq b\rq\rq, the \fermilat\ light curve of \pks\ is reported (starting from February 13, 2015), using a time binning of 6\ h around the \gray\ flare (MJD\,57070--57074) and 12\ h elsewhere, and a power-law spectrum with both the flux and photon index of \pks\ free to vary. In the light curve, flux upper-limits at the 68$\%$ level are provided when $TS <9$. The evolution in time of the best-fit photon index is shown in panel \lq\lq c\rq\rq.\\

To better compare the \fermilat\ data of \pks\ with \hess\ data during the night it was detected at VHE, as well as to retain sufficient statistics, a \fermilat\ data subset was analyzed for the time window MJD\,57072.5--57073.5. The source is detected in this subset with a TS=339 ($\sim$18$\sigma$), a flux of $F_{(100\,\rm{MeV} - 500\,\rm{GeV})}=(2.08 \pm 0.26) \times 10^{-6}$\,cm$^{-2}$\,s$^{-1}$, and a photon index of $\Gamma=2.15 \pm 0.10$. In this case, no evidence of curvature is found (log-likelihood ratio of -0.12). The \fermilat\ highest energy for a photon associated with \pks\ (at the  95$\%$ confidence level) is 106 GeV (detected on MJD 57072, at 14:42:20 UTC). This is also the highest photon energy when considering the larger two-months interval described above. Given the high significance of the detection, a \fermilat\ data analysis has also been performed in a time window strictly simultaneous with the \hess\ detection (MJD\,57072, 18:53-20:55 UTC). \pks\ is detected in this dataset with a TS= 37 ($\sim$6$\sigma$), a flux of $F_{(100\,\rm{MeV} - 500\,\rm{GeV})}=(4.57 \pm 1.52) \times 10^{-6}$\,cm$^{-2}$\,s$^{-1}$, and a photon index of $\Gamma=2.43 \pm 0.33$. The highest-energy LAT photon simultaneous with the \hess\ detection has $E=1.71$ GeV (detected on MJD 57072, at 20:40:16 UTC). Fig. \ref{gammaspectra} shows the \fermilat\ spectra of \pks\ for the two time windows described here.\\

To estimate the variability in the HE \gray\ band, the fractional variability $F_{var, HE}$ of the 12-hours binned light curve was computed following \citet{vaughan03}. This results in $F_{var, HE} = (62 \pm 7)$\%, clearly supporting the presence of variability of the HE \gray\ flux. This finding does not depend on the binning of the light curve: for a 6-hours binned light curve, $F_{var, HE} = (48 \pm 2)$\%, while for a 24-hours binned light curve,  $F_{var, HE} = (57 \pm 4)$\%. An important property of a flare is its variability timescale, which is defined here as the flux-doubling timescale \citep[see e.g.][]{veritasmrk5012009}:
\begin{center}
\begin{equation}
F(t) = \frac{a}{2^{-(t-t_0)/b} + 2^{(t-t_0)/c}}
\end{equation}
\end{center}
where $t_0$ is the time of the peak flux in the light curve, $a$ fits the peak flux, and $b$ and $c$ denote the rising and falling flux-doubling time-scales, respectively.
 The evolution of the \gray\ flare in the HE band shows a slow flux increase starting on February 13, 2015, with a first flare on February 18, 2015, at an integral flux level of around $3\times10^{-6}$ cm$^{-2}$ s$^{-1}$. The flux then dropped by a factor of around four, before brightening again on February 19, 2015, coincidentally with the \hess\ detection. The flux doubling time-scale for this second flare centered on the \hess\ detection is fitted using the 6-hours binned light curve as $6 \pm 4$ hours for the rising and $4 \pm 2$ hours for the falling part of the flare. \\

\subsection{Swift-XRT}

Following the detection of the \gray\ flaring activity in \pks, ToO observations were requested to the \textit{Neil Gehrels} \swift\ satellite \citep{swift} in order to measure the flux of \pks\ in soft-X-rays and optical/UV. The ToO resulted in four observations with \swift\ starting from February 20, 2015, for a  total exposure time of around 12.3 ks. The details are provided in Table \ref{Tableswift}.\\

Data from \xrt\ \citep{swiftxrt} are analyzed using \textit{Heasoft} version 6.18. All observations are taken in the standard \textit{Photon Counting} observing mode. The event files are cleaned using standard screening criteria. Images, spectra and light curves are extracted using a circular region with radius equal to 20 pixels. The background is extracted from an annular region with inner radius equal to 50 pixels, and outer radius equal to 160 pixels. The count rates are around  0.11-0.16 s$^{-1}$, and are low enough that no pile-up affects the analysis.\\

Spectral analyses of \xrt\ data are performed using \textit{XSpec} version 12.9. The ancillary response files are recomputed using \textit{xrtmkarf}, while the redistribution matrix files provided by the \swift\ team are used. Data are rebinned using \textit{grppha} imposing a minimum of 30 counts per bin, and data below 0.3 keV are excluded.\\

 The X-ray spectrum from each of the four \xrt\ observations is fitted with an absorbed powerlaw model (\textit{phabs*cflux*powerlaw} within \textit{Xspec}, to access the unabsorbed flux in the 0.3-10 keV band), to take into account the absorption by neutral material in the Milky Way. The value of $N_H$ in the direction of \pks\ is fixed to $7.8 \times 10^{20}$ cm$^{-2}$ as estimated by \citet{Dickey90}. The results of the fit are provided in Table \ref{Tableswift}. The \xrt\ light curve is shown in Fig. \ref{LCMWL}, panel \lq\lq d\rq\rq. No significant variability is detected in the X-ray data: the four \xrt\ observations are consistent with a constant flux ($\chi^2 / DOF = 9/3$, which corresponds to a chance probability of about 3$\%$), with an average $F_{0.3-10\ \textrm{keV}}$ = (8.1 $\pm$ 0.3)\ 10$^{-12}$ erg cm$^{-2}$ s$^{-1}$.\\

\subsection{Swift-UVOT}

Simultaneously with \xrt, the UVOT telescope \citep{swiftuvot} also observed \pks, guaranteeing a coverage in the optical and UV band. During the first two observations all UVOT filters were available, while the two other observations were taken only with the W1 and U filter, respectively (see Table \ref{Tableswift}). Fluxes are calculated using \textit{uvotmaghist}, integrating over a 5\arcsec-radius circular region for the source, and a 20\arcsec-radius circular region for the background. Data are de-reddened following \citet{uvotgrb} using $E_{B-V}=0.121$ as value for the Galactic extinction \citep{schlafly11}.
The \uvot\ light curve is presented in Fig. \ref{LCMWL}, panel \lq\lq e\rq\rq. There is clear variability in the optical band, showing a decrease from February 20, 2015 to February 27, 2015. \\

\subsection{ATOM}

The ATOM telescope \citep{atom}, located on the \hess\ site, regularly observes \pks\ as part of its blazar monitoring program, providing a long term light curve of the source in the R band. Fluxes from ATOM observations are extracted using a 4\arcsec-radius aperture, and have been de-reddened using the same value of Galactic extinction $E_{B-V}=0.121$ as for the UVOT data analysis. There are no ATOM observations during the night of February 19, 2015, when \hess\ detected VHE emission from \pks. On the other hand, ATOM observations on the night of February 20, 2015, are contemporaneous with the UVOT ones. The light curve from ATOM data is provided in Fig. \ref{LCMWL}, panel \lq\lq e\rq\rq. \\

\begin{figure}[t!]
\begin{center}
\begin{tikzpicture}
 \node[anchor=south west,inner sep=0] (image) at (0,0) {\includegraphics[width=260pt]{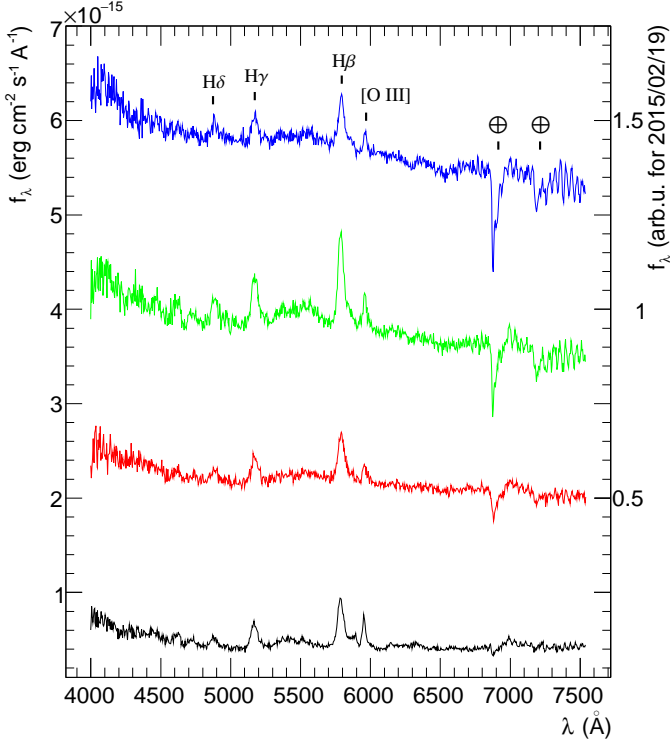}};
  \begin{scope}[x={(image.south east)},y={(image.north west)}]
                \draw[black,thick] (0.505,0.83) -- (0.505,0.84);
                 \draw[black,thick] (0.54,0.785) -- (0.54,0.795);
                 \draw[black,thick] (0.32,0.805) -- (0.32,0.815);
                 \draw[black,thick] (0.38,0.81) -- (0.38,0.82);
                 \draw[black,thick] (0.73,0.75) -- (0.73,0.76);
                    \draw[black,thick] (0.79,0.75) -- (0.79,0.76);
    \end{scope} 
    \node[scale=0.8] at (2.94,9.1) {H$\delta$};
    \node[scale=0.8] at (3.5,9.15) {H$\gamma$};
    \node[scale=0.8] at (4.65, 9.35) {H$\beta$};
    \node[scale=0.8] at (5.2, 8.9) {[O III]};
     \node[scale=1] at (6.68, 8.52) {$\Earth$};
     \node[scale=1] at (7.24, 8.52) {$\Earth$};
       \node[scale=0.7] at (7.98, 0.68) {$^\circ$};
\end{tikzpicture}
\caption{Optical spectra (4000-7500 $\AA$) of \pks\ obtained with SPOL on three consecutive nights of 2015: February 18 (in red), February 19 (in green), and February 20 (in blue). Due to absence of photometric conditions on February 19, this spectrum is provided in arbitrary units. For reference, it is shown in black a spectrum from a low-state of the source (February 9, 2016). The emission lines and telluric features are identified in the spectrum from February 20, 2015. \label{sospectra}}
\end{center}
\end{figure} 

\subsection{Steward Observatory}

\pks\ is among the sources covered by the Steward Observatory blazar monitoring program \citep{steward, carnerero15}. The program provides optical monitoring for bright \fermilat\ blazars, including photometry, spectroscopy and spectro-polarimetry. During February 2015, \pks\ was observed on nine
occasions from February 12 to February 20, using the 1.5-m Kuiper Telescope located on Mt. Bigelow, Arizona
and the SPOL CCD spectropolarimeter \citep{schmidt92}. The SPOL apertures range were 5.1\arcsec $\times$ 10\arcsec with the first number being the width of the slit and the second number being the width of the spectroscopic extraction aperture.  The extraction aperture is in the east-west direction on the sky.  For the SPOL differential photometry measurements, the slit width used was 12.7\arcsec and the extraction aperture was either 10\arcsec, 11\arcsec, or 12\arcsec, depending on the seeing at the time of the measurement. Photometry measurements are provided in the V (Johnson) and R (Kron-Cousins) bands, and have been de-reddened using $E_{B-V} = 0.121$ as for the UVOT and ATOM data. The light curve from the Steward Observatory monitoring of \pks\ is shown in Fig. \ref{LCMWL}, and complements UVOT and ATOM measurements to provide optical coverage of the \gray\ flaring activity in \pks. 
Observations using SPOL were made on the morning of February 20, 2015 UTC, 8.4 hours after the \hess\ detection, and
show a clear optical flare. There was an increase in luminosity of about one magnitude compared to the measurements taken 48 hours previously with SPOL and with the Swift-UVOT and ATOM observations made 17 hours afterwards. Measurements of the equivalent widths of the H$\beta$ and H$\gamma$ broad-emission lines of \pks\ using the flux spectra obtained with SPOL confirm the rapid brightening of the blazar on February 20.  The decrease in the equivalent widths of the emission lines from February 19 to February 20 is
consistent with a 1-mag increase in the continuum brightness over the period of about 24 hours. The onset of the optical flare appears to be as rapid as its decay (see panel \lq\lq e\rq\rq\  in Fig. \ref{LCMWL}), and the coincidence of the optical, HE, and VHE outbursts is strong evidence that they are closely related physically.\\

Measurements of the optical polarization fraction and position angle are shown in panels \lq\lq f\rq\rq\ and \lq\lq g\rq\rq\ of  Fig. \ref{LCMWL}. Uncertainties on both quantities are smaller than the symbols used: the uncertainty on the polarization fraction is $\leq 0.2\%$, while the uncertainty on the angle is $\leq1^\circ$. The polarization fraction reported has been corrected by subtracting the assumed unpolarized contributions to the spectrum by the BLR and big blue bump.  This is best done by subtracting an optical spectrum of
PKS0736+017 obtained during a faint period when the blazar shows little polarization.  This is when the BLR and thermal optical continuum are most dominant in the spectrum.  After subtracting the unpolarized flux, the measured polarization fraction is a better approximation to the intrinsic polarization of the synchrotron continuum responsible for the rapid optical flux variations.   Given that the spectrum of PKS0736+017 still includes some polarization from the non-thermal continuum even when it is optically faint, the values of the polarization fraction plotted in Figure \ref{LCMWL} are the highest possible intrinsic polarization levels of the synchrotron emission. The highest polarization is observed not for the night of the optical flare, but for the night before, at $(12.5 \pm 0.1) \%$. The polarization position angle is extremely variable during the week leading up to the \gray\ flare and is observed throughout the full 180$^\circ$ range. It is interesting to note that the two measurements taken on February 20 show a variation of the polarization angle of $(44.9 \pm 0.6)^\circ$ in three hours, which is a very rapid rotation even for a blazar \citep[see][]{blinov15, blinov16, lyutikov17}.\\
 
Spectro-polarimetry of \pks\ yields the spectral index $\alpha$ (defined as $f_\lambda\ \propto\ \lambda^{-\alpha}$) of the optical synchrotron continuum during the flare, which is key to constraining radiative models (see Section \ref{sedsection}). The spectral index is derived directly from the polarized flux spectrum, assuming that the polarization of the synchrotron continuum is constant with wavelength. This may not be the case, but observations of BL Lac objects, where the synchrotron emission dominates the optical flux, do not often show strong wavelength-dependent polarization \citep{Sitko85, Jannuzi94, Smith96}. Also, PKS0736+017 shows no evidence for a strong variation in polarization position angle with wavelength in 2015 February. The de-reddened polarized flux spectrum of \pks\ is consistent with a featureless power law and indicates that the emission lines are unpolarized.  During the morning of Feb 20, 2015, the spectral index is found to be $1.0 \pm 0.1$, significantly flatter than for the nights before the flare, when it is observed to be between $1.4$ and $1.8$, indicating a shift of the synchrotron peak frequency to higher energies.\\ 
 
\subsection{ASAS-SN and KAIT} 

\pks\ is also regularly monitored in the optical by the ASAS-SN project \citep{asassn} and the KAIT telescope \citep{cohen}. Their public data have been retrieved for the observations around the February 2015 \gray\ flare. ASAS-SN V-magnitudes, and KAIT unfiltered magnitudes \citep[which are close to the R ones, see][]{li03}  have been converted to fluxes and de-reddened using $E_{B-V} = 0.121$ as for the other optical observations. ASAS-SN and KAIT observed \pks\ during the morning (UTC) of February 19, respectively 11.5 hours and 14 hours before the \hess\ VHE detection, and complement the optical light curve from UVOT, ATOM, and Steward Observatory. The contribution of the host-galaxy to the optical measurements has not been investigated. The emission in the optical band is dominated by the non-thermal continuum from the quasar, as it is evident from the SED shown in Fig. \ref{sed}.\\

\section{Discussion}
\label{section3}

\subsection{On the location of the \gray\ emission region}

As shown in the multi-wavelength light curve in Fig. \ref{LCMWL}, the only instrument which was observing \pks\ together with \hess\ during the night of February 19, 2015, was \fermilat. Observations by \swift\ covered only the post-flare period, and there are no optical observations available within 8 hours of the \hess\ detection. It is thus not possible to follow the usual approach of modeling the simultaneous SED to access blazar physics. In particular, there is no coverage of the behavior of the synchrotron component of \pks\ simultaneously with the \gray\ flare.  On the other hand, constraints can be put on the location of the emitting region $r$ (defined as the distance from the SMBH) within the relativistic jet, under some hypotheses. In the following, it is assumed that\footnote{Here and in the following, quantities in the co-moving jet frame are indicated by a prime.}:
\begin{itemize}
\item the main radiation mechanism responsible for the \gray\ emission is inverse-Compton scattering by a population of electrons/positrons in the jet, off a low-energy photon field represented by the radiation from the accretion disc, the BLR and the dusty torus; 
\item the \gray\ emission during the flare is produced from a single region within the relativistic jet;
\item the jet of \pks\ is closely aligned to the line of sight, and the Doppler factor $\delta$ of the emitting region equals its bulk Lorentz factor $\Gamma$, that is the angle to the line of sight $\vartheta_{los}$ is equal to $1/\Gamma$;
\item the emitting region is approximated by a spherical plasmoid in the jet, characterized by its radius $R^\prime$, which is related to the variability time-scale $\tau_{var}$ via $R^\prime \simeq c\ \tau_{var}  \frac{\Gamma}{1+z}$, where $c$ is the speed of light in vacuum, and $z$ is the redshift of the source; the most constraining estimate for $\tau_{var}$ comes from the falling part of the \fermilat\ flare as $\tau_{var} = (4 \pm 2)$ hours;  the two extreme values of 2 and 6 hours are used in the following;
\item the emitting region fills the entire cross-section of the relativistic jet and thus the location of the emitting region $r$ and its size $R$ are simply related by $R/r= \tan \vartheta_{open}$, where $\vartheta_{open}$ is the jet opening angle;
\item the location of the BLR, $r_{BLR}$, is derived from the luminosity of the H$\beta$ line, which is measured from Steward Observatory data to be $L_{H\beta} = 4.20 \times 10^{42}\ \textrm{erg s}^{-1}$. Following \citet{greene05}, $r_{BLR}$ is then estimated as $1.45 \times 10^{17}$ cm; the total BLR luminosity is similarly estimated from $L_{H\beta}$ \citep[see][]{finke16} as $L_{BLR} =  1.24 \times 10^{44} \textrm{erg s}^{-1}$, and  $L_{disk} \simeq 10\ L_{BLR} = 1.24 \times 10^{45} \textrm{erg s}^{-1}$. The location of the dusty torus $r_{torus}$ is assumed to scale as $r_{torus} \simeq 2.5\times 10^{18} \sqrt{ L_{disk} / 10^{45} \textrm{erg s}^{-1}} = 2.85 \times 10^{18}$ cm \citep[see e.g.][]{sikora09, hayashida12}. The values for the luminosity of the accretion disc and the SMBH mass of \pks\ vary largely in the literature: estimates for $L_{disk}$ range from $10^{44.6}$ to $10^{45.7} \textrm{erg s}^{-1}$, while estimates for $M_\bullet$ range from $10^8$ to $10^{8.7}\ M_\odot$  \citep{wandel91, mclure01, woo02, Marchesini04, dai07}. The adopted value of $L_{disk} \simeq 1.24 \times 10^{45} \textrm{erg s}^{-1}$ is consistent with these previous estimates;
\item the BLR is modeled as a spherical shell centered at $r_{BLR}$, with lower boundary $r_{in} = 0.9*r_{BLR}$ and outer boundary $r_{out} = 1.1*r_{BLR}$; as discussed in \citet{bottcheropacity}, the choice of the boundaries have negligible effects on the BLR opacity; the opening angle of the dusty torus is assumed to be $\pi/4$ as in \citet{nalew14}.

\end{itemize}

\begin{figure}[t!]
\begin{center}

\begin{tikzpicture}
    \node[anchor=south west,inner sep=0] (image) at (0,0) {\includegraphics[width=260pt]{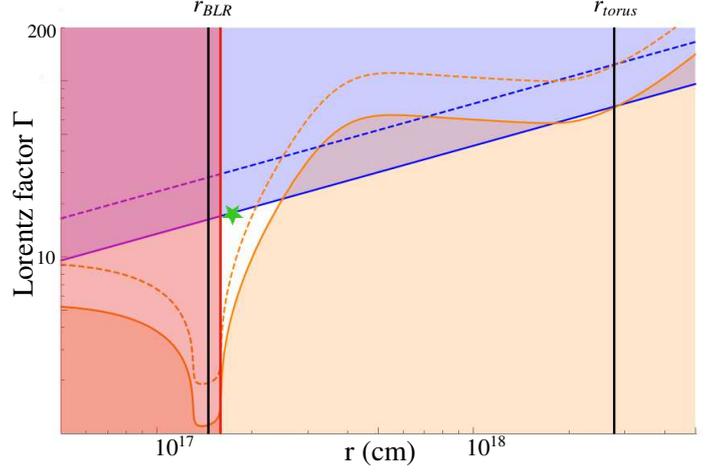}};
    \node[scale=1] at (2.3,5.9) {$r_{BLR}$};
    \node[scale=1] at (7.6,5.9) {$r_{torus}$};
    \node[scale=1.2] at (4.5,0.) {r (cm)};
     \node[scale=1.2, rotate=90] at (-0.2,3.3) {Lorentz factor $\Gamma$};
      \node[scale=1] at (1.8,0.1) {$10^{17}$};
       \node[scale=1] at (5.9,0.1) {$10^{18}$};
        \node[scale=0.8] at (0.1,2.6) {10};
     \node[scale=0.8] at (0.05,5.6) {200};
\end{tikzpicture}
\caption{Constraint on the location of the emitting region $r$ in cm, defined as distance from the SMBH, as a function of the bulk Lorentz factor $\Gamma$.  The red exclusion region represents the opacity constraint, the blue exclusion region represents
the collimation constraint, while the orange exclusion region represents the cooling constraint, the last two computed for $\tau_{var} = 6$ hours. Dashed lines show how the constraints change assuming $\tau_{var} = 2$ hours. The white regions indicate the allowed part of the parameter space. The vertical
black lines represents the estimated locations of the BLR $r_{BLR}$ and of the dust torus $r_{torus}$. The green star identifies the parameters of the EIC solution discussed in Sec. \ref{sedsection} and plotted in Fig. \ref{sed}. \label{constraintsregion}}
\end{center}
\end{figure}

The first constraint on the location of the \gray\ emitting region comes from opacity to $\gamma-\gamma$ pair-production. The VHE photons can pair-produce over the bright environment of the SMBH and, as shown by several authors \citep[see e.g.][]{donea03,liubai, reimer07,tavecchio12, bottcheropacity, finke16}, the absorption can be so severe that the simple detection of VHE photons can be used to exclude an emitting region located at the base of the jet. The \gray\ spectra presented in Fig. \ref{gammaspectra} show that the extrapolated \fermilat\ spectrum, once the absorption on the EBL is taken into account, is consistent with the \hess\ detection. This translates into an internal opacity equal to zero, constraining the location of the emitting region well beyond $r_{BLR}$. However, given the large uncertainty in the simultaneous \fermilat\ spectrum, the most conservative opacity constraint has to be computed using the upper-end of the \fermilat\ extrapolated bow-tie. In this case a break exists between the \fermilat\ and \hess\ energy band. This break can be intrinsic (for example due to a break in the lepton population, or due to the transition from the Thomson to the Klein-Nishina regime of the inverse-Compton scattering), or due to additional pair-production absorption at the source. In particular, while pair-production on the infrared photons from the dusty torus is expected to produce a spectral break at a few TeV, absorption on Ly$\alpha$ photons can produce a spectral break at around 100 GeV. Given that it is not possible to discriminate between an intrinsic cut-off in the \gray\ emission and an absorption effect, the opacity argument can only be used to put a lower limit on the location of the emitting region $r_{min}$: if the emitting region is located further in, closer to the SMBH, the spectral break between the \fermilat\ and \hess\ observations would have been stronger. To quantitatively estimate $r_{min}$, the model described in \citet{bottcheropacity} is used, together with the values of $L_{disk}$ and $r_{BLR}$ provided above. The calculation of $r_{min}$ is performed by extrapolating into the VHE band the upper-end of the \fermilat\ bow-tie, including absorption by internal and EBL photons, and varying $r$ until the extrapolated spectrum matches the \hess\ measurement. This estimate results in $r_{min} = 1.6\times 10^{17}$ cm, or $1.1\ r_{BLR}$. This limit is shown in Fig. \ref{constraintsregion} by the red exclusion region.\\

The second constraint comes from the collimation of the relativistic jet. As shown by radio observations, relativistic jets from SMBHs are highly collimated and in particular $\Gamma\ \vartheta_{open} <1$. In the following, the reformulation by \cite{nalew14} is used, which translates this inequality into a limit in the $\Gamma - r$ plane, under the assumption that $R^\prime \simeq c\ \tau_{var}  \frac{\Gamma}{1+z}$, and that $R^\prime/r= \tan \vartheta_{open}$. This constraint is shown in Fig. \ref{constraintsregion} by the blue exclusion region.\\

The last constraint comes from the cooling of the leptons due to inverse Compton scattering. In particular, the cooling time-scale $\tau_{cool}$ is required to be shorter than the observed variability time-scale $\tau_{var}$, under the assumption again that $R^\prime \simeq c\ \tau_{var}  \frac{\Gamma}{1+z}$, and that the cooling time-scale is dominated by the inverse Compton scattering on the external photon fields. To translate this condition into an exclusion region in the $\Gamma - r$ plane, it is used a modified version of the equations described in \citet{nalew14}: instead of considering a BLR opening angle of $\pi / 4$, as assumed in the original paper,  the equations have been recomputed for a spherical BLR, to be consistent with the opacity constraint. The impact of this change is to open up the parameter space, allowing a larger range of $\Gamma$ values when the emitting region is located close to $r_{BLR}$. The cooling constraint is shown in Fig. \ref{constraintsregion} by the orange exclusion region. The cooling time-scale depends on the energy density of the target photons, and thus depends on $r$: the shape of the exclusion region in Fig. \ref{constraintsregion} is due to the changes in the photon field seen by the \gray\ emitting region when approaching the BLR radius $r_{BLR}$ and torus radius $r_{torus}$.\\

The three aforementioned constraints limit significantly the $\Gamma - r$ plane. A location of the emitting region close to the SMBH, where the dominant photon field is the thermal radiation from the accretion disc, is excluded due to the opacity constraint. Two scenarios are, however, allowed: an emitting region located at around $r_{BLR}$, with $\Gamma \simeq 10-20$, or an emitting region close to $ r_{torus}$, with $\Gamma \simeq 60$. In the first scenario, the dominant external photon field is the emission from the BLR, while in the second case it is the thermal emission from the dusty torus. The first solution implies lower values of $\Gamma$, in line with radio observations of \gray\ FSRQs and of \pks\ in particular, for which an estimate of $\Gamma = 16.5 - 17.0$ is provided by \citet{push09} and \citet{hovatta09}, respectively, from direct measurements of the jet speed. Changing the assumed value of $\tau_{var}$ results in a translation of the constraints towards larger values of $\Gamma$, as shown in Fig. \ref{constraintsregion}. \\

\subsection{Spectral Energy Distribution}
\label{sedsection}

\begin{figure*}[t!]
\begin{center}
\includegraphics[width=360pt]{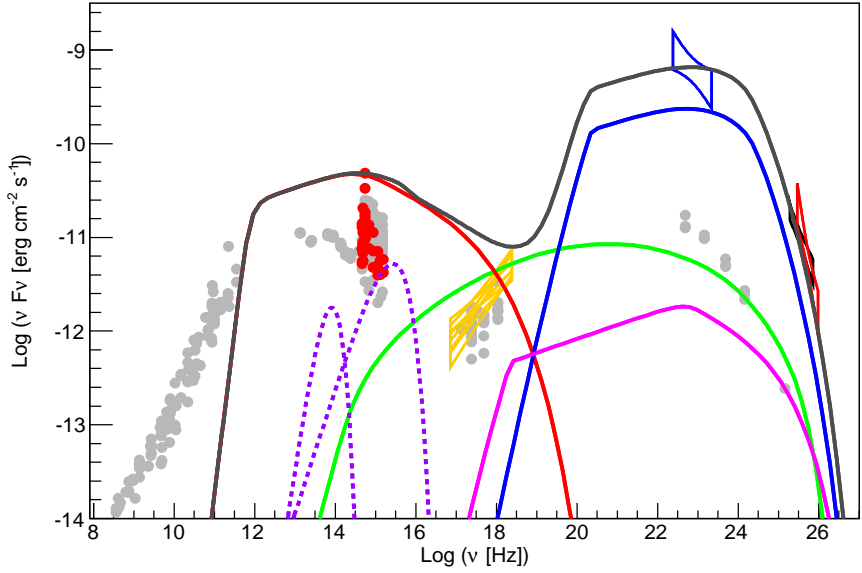}
\caption{Spectral energy distribution of \pks\ during the multi-wavelength campaign of February 2015, together with archival data (in grey). The black and red bow-ties represent the \hess\ spectra during the night of February 19, 2015, for the monoscopic and stereoscopic reconstruction, respectively; the blue bow-tie represents the \fermilat\ spectrum simultaneous with the \hess\ detection; the orange bow-ties represents the four Swift-XRT spectra, acquired after the \hess\ detection; the red points represent all optical/UV measurements before and after the flare, from ATOM, Steward Observatory and Swift-UVOT. The EIC model components, from low to high energies are: synchrotron emission by leptons (in red), synchrotron-self-Compton emission (in green), EIC emission over Ly$\alpha$ line (in blue, emission from EIC over other BLR lines is not plotted for clarity purposes), EIC emission over the dust torus (in magenta). The violet dotted lines represent the thermal emission from the dusty torus, and the accretion disk, respectively. \label{sed}}
\end{center}
\end{figure*}

The SED of \pks\ during the February 2015 flaring activity, using the data described in the previous section and together with archival observations\footnote{from the SSDC SED builder, \url{https://tools.ssdc.asi.it}.}, is presented in Fig. \ref{sed}. As discussed previously, it is not possible to fit the overall SED during the night of the \hess\ detection due to the absence of simultaneous information on the behavior of the synchrotron component.  It is however interesting to use the constraints on $\Gamma$ and $r$ computed in the previous section as input parameters for a leptonic model fitting the \gray\ SED. The goal is first to see if it is possible to find a model which remains consistent with the observations during the flare, and for what parameter values (other than $\Gamma$ and $r$), and then discuss the implications for the behaviour of the synchrotron component during the flare.
For this purpose, the \gray\ emitting region is assumed to be at $r =1.7\times10^{17}$ cm, and located at the limit of the collimation constraint, i.e. $\Gamma\ \vartheta_{open} = 1$ (the solution is identified as a green star in Fig. \ref{constraintsregion}).  The Lorentz factor $\Gamma$ is equal to $17.7$, and the size of the emitting region $R$ is estimated via $\tau_{var} = 6$ hours to be $9.6\times 10^{15}$ cm. The ratio of $R/r$ gives a jet opening angle $\vartheta_{open}$ of $3.2^\circ$ (which by construction respects $\Gamma\ \vartheta_{open} = 1$). The estimate of the jet opening angle of \pks\ from radio observations is smaller, $\vartheta_{open} = 1.8^\circ \pm 0.3^\circ$ \citep{pushk17}. This value refers however to the jet opening angle as measured at much larger (kpc) distances from the SMBH. The fact that the intrinsic opening angle gets smaller downstream is consistent with the findings of \citet{pushk17} from an analysis of 65 AGNs. The leptons in the emitting region scatter primarily BLR photons: the BLR photon energy density at $r$ is estimated following \citet{nalew14}  resulting in $u_{BLR} = 3.95$ erg cm$^{-3}$. The energy density of torus photons is similarly estimated as $u_{torus} = 0.005$ erg cm$^{-3}$, clearly negligible with respect to the BLR one. The numerical code used to compute synchrotron and EIC radiation is described in \citet{Cerruti13}. The electron distribution in the emitting region is modeled by a broken power-law function with exponential cut-off 
\begin{equation}
  N^\prime_e(\gamma^\prime) = \begin{cases}
        N^\prime_0  \gamma^{\prime -n_1} & \textrm{if}\  \gamma^\prime_{min} \leq  \gamma^\prime \leq \gamma^\prime_{break}
        \\
        N^\prime_0 \gamma^{\prime (n_2-n_1)}_{break}   \gamma^{\prime -n_2} e^{-\gamma^\prime/\gamma^\prime_{Max}} & \textrm{if}\  \gamma^\prime_{break} <  \gamma^\prime 
        \end{cases}
 \end{equation}

with indices $n_1$ and $n_2$ below and above the break Lorentz factor $\gamma^\prime_{break}$, cut-off Lorentz factor $\gamma^\prime_{Max}$, and minimum Lorentz factor $\gamma^\prime_{min} = 10$. A complex BLR line spectrum is assumed, using the seven most prominent emission lines from the average quasar model by \citet{telfer02}. The absorption on BLR photons is calculated following \citet{dermer09}, under the assumption that the most relevant absorption is the one due to the Ly$\alpha$ line. Absorption by the EBL is computed using the model by \citet{franceschini08}. The remaining free parameters are adjusted to reproduce the \gray\ data from \fermilat\ and \hess, a peak of the synchrotron component $\nu_{sync} = 8\times\ 10^{14}$~Hz, as constrained by the flat optical spectrum measured by Steward Observatory observations 8 hours after the \hess\ detection, and a flux of the synchrotron peak similar to the maximum observed optical flux. A good description of the SED can be obtained assuming $n_1=2.7$, $n_2=3.4$, $\gamma_{break}=2\times10^3$, $\gamma_{Max}=1\times10^5$, $N_0^\prime=1\times10^6$ cm$^{-3}$, and a magnetic field $B^\prime =1.6$ G. The corresponding value of the equipartition factor (the ratio of the electron energy density to the magnetic energy density) $u^\prime_e/u^\prime_B$ is $2.3$. The resulting modeling is shown in Fig. \ref{sed}. The peak of the EIC component of the model is in the GeV range, resulting in a flat spectrum in the LAT energy band. The hard HE spectrum is needed in order to fit the VHE spectrum, as both the transition to the Klein-Nishina regime and the absorption on BLR photons lead to an under-prediction of the VHE spectrum for softer HE spectra. The model indicates that an X-ray flare could have occurred simultaneously with the VHE \gray\ flare, with significant softening of the spectrum, due to the emergence of the synchrotron component, similar to the one observed in PKS 1441+25 \citep{1441veritas, 1441magic}. Although the model predicts a soft-X-ray flare a factor of about ten brighter at 0.1 keV compared to the \xrt\ observations of \pks, it is important to note that the electron synchrotron cooling time-scale at these energies is very short: for electrons with Lorentz factor $\gamma=10^{4}$  in a magnetic field $B^\prime = 1.6$ G, and moving towards the observer with bulk Lorentz factor $\Gamma = 17.7$, the synchrotron cooling time-scale is only 35 minutes, to be compared with the 24 hours delay of the \xrt\ observations with respect to the time of the \hess\ detection. This prediction of an unseen soft-X-ray flare is solid with respect to the free parameters considered: having fixed $\Gamma$, $R$, and $u_{BLR}$, the normalization of the particle distribution $N^\prime_0$ and the magnetic field $B^\prime$ are adjusted to reproduce the peak flux of the synchrotron and EIC components, and $\gamma^\prime_{break}$ is adjusted to have a synchrotron peak in the optical band. The high value of $\gamma^\prime_{Max}$, which is the parameter that implies the occurrence of a simultaneous X-ray flare from the source, is required for the model to reach VHE \gray s.\\

\section{Conclusions}

\hess\ observations of the FSRQ \pks, triggered on the basis of a \gray\ flare detected with \fermilat, resulted in the discovery of VHE emission from this quasar during the night of February 19, 2015. \pks\ is the seventh member of the elusive population of FSRQs known to emit VHE photons, the nearest so far. \fermilat\ and \hess\ high-energy flares were accompanied by at least a 1 mag brightening of the non-thermal optical continuum.
  \fermilat\ observations show the presence of a relatively fast \gray\ flare, with flux-doubling time-scale of around six hours. No optical nor X-ray observations were performed strictly simultaneously with the \hess\ detection. Nonetheless, both temporal and spectral measurements in the \gray\ band can be used to put model-dependent constraints on the location of the \gray\ emitting region in the jet, and on its Lorentz factor. The \fermilat\ and the \hess\ spectra collectively constrain the location of the emitting region to be located beyond $r = 1.1\ r_{BLR} = 1.6\times 10^{17}$ cm to avoid  absorption due to $\gamma$-$\gamma$ pair-production with BLR photons. A location of the \gray\ emitting region just outside the BLR with a Lorentz factor $\Gamma \simeq 10-20$ is thus a viable solution which satisfies both temporal and spectral constraints. Alternatively, the emitting region may be located much farther away, at around $r_{torus} = 2.85 \times 10^{18}$ cm, with a higher value of $\Gamma \simeq 60$, with electrons in the jet in this case scattering thermal photons from the dusty torus.\\

The main limitation of this study is the absence of strictly simultaneous observations of the behavior of the synchrotron component during the \hess\ detection. With simultaneous optical, UV, and X-ray data, it will be possible to uniquely constrain the electron energy distribution, and the opacity and variability constraints could be coupled with a full SED fitting. In addition with tighter constraints on the location of the \gray\ emitting region, it will be possible to study in more details the opacity in the VHE band, potentially putting constraints on the geometry of the BLR itself (its aperture angle and its profile). This kind of study is more easily done using nearby quasars, for which the internal absorption is not hidden by the EBL absorption. In this context, \pks\ represents the ideal FSRQ, being located much closer than all other VHE FSRQs. Its location near the celestial equator makes it also a perfect target for all current ground-based \gray\ telescopes, with only marginal zenith-angle effects. The next \gray\ flare from \pks\ has thus the potential to be an extremely interesting event for the study of \gray\ quasars, and future multi-wavelength coordinated observing campaigns are strongly encouraged.\\       

\label{section4}
\begin{acknowledgements}
The support of the Namibian authorities and of the University of Namibia in facilitating 
the construction and operation of H.E.S.S. is gratefully acknowledged, as is the support 
by the German Ministry for Education and Research (BMBF), the Max Planck Society, the 
German Research Foundation (DFG), the Helmholtz Association, the Alexander von Humboldt Foundation, 
the French Ministry of Higher Education, Research and Innovation, the Centre National de la 
Recherche Scientifique (CNRS/IN2P3 and CNRS/INSU), the Commissariat à l’énergie atomique 
et aux énergies alternatives (CEA), the U.K. Science and Technology Facilities Council (STFC), 
the Knut and Alice Wallenberg Foundation, the National Science Centre, Poland grant no. 2016/22/M/ST9/00382, 
the South African Department of Science and Technology and National Research Foundation, the 
University of Namibia, the National Commission on Research, Science $\&$ Technology of Namibia (NCRST), 
the Austrian Federal Ministry of Education, Science and Research and the Austrian Science Fund (FWF), 
the Australian Research Council (ARC), the Japan Society for the Promotion of Science and by the 
University of Amsterdam. We appreciate the excellent work of the technical support staff in Berlin, 
Zeuthen, Heidelberg, Palaiseau, Paris, Saclay, Tübingen and in Namibia in the construction and 
operation of the equipment. This work benefited from services provided by the H.E.S.S. 
Virtual Organisation, supported by the national resource providers of the EGI Federation.  The CC-IN2P3 (\href{https://cc.in2p3.fr}{cc.in2p3.fr}) is gratefully acknowledged for providing a significant amount of the computing resources and services needed for this work. Part of this work is based on archival data, software or online services provided by the Space Science Data Center - ASI. Blazar observations at Steward Observatory are funded though NASA/Fermi Guest Investigator Program grant NNX15AU81G. Matteo Cerruti has received financial support through the Postdoctoral Junior Leader Fellowship Programme from la Caixa Banking Foundation (LCF/BQ/LI18/11630012). The authors would like to thank the anonymous referee, whose constructive comments significantly improved the manuscript. 
\end{acknowledgements}

\bibliographystyle{aa}
\bibliography{PKS0736_biblio}

\end{document}